\documentclass[prl, twocolumn]{revtex4}
\usepackage{epsfig}

\begin{document}
\title{KLM quantum computation as a measurement based computation}

\author{Sandu Popescu$^{a,b}$}
\affiliation{$^c$H. H. Wills Physics Laboratory, University of
Bristol, Tyndall Avenue, Bristol BS8 1TL}
\affiliation{$^b$Hewlett-Packard Laboratories, Stoke Gifford, Bristol
BS12 6QZ, UK}
\date{\today}

\begin{abstract}
 We show that the Knill Laflamme Milburn method of quantum computation with linear optics gates can be interpreted as
 a one-way, measurement based quantum computation of the type introduced by Briegel and Rausendorf. We also show that
 the permanent state  of $n$ $n$-dimensional systems is a universal state for quantum computation.

\end{abstract}


\maketitle

\newcommand{\beq}{\begin{equation}}
\newcommand{\eeq}{\end{equation}}
\newcommand{\ra}{\rangle}
\newcommand{\la}{\langle}
\newcommand{\ket}[1]{\left| #1 \right\rangle}

The first model for quantum computation,  the ``gate array" model, resembles very closely
 the way classical computers are constructed. In a classical computer the bits are input into gates which perform
elementary logic operations on them, such as NOT, AND and OR; the bits are then passed from one gate to another.
A quantum gate array is very similar with the difference that it uses qubits instead of bits and that the gates
     perform elementary unitary transformations instead of ordinary logic operations.  Recently however, alternative
      models for quantum computation were suggested, such as adiabatic quantum computation \cite{adiabatic},
      topological computation \cite{topological}, one-way  measurement based computation (cluster states)
      \cite{oneway} and the Knill, Laflamme Milburn (KLM) linear optics computation \cite{klm} (which relies on
      the Bose-Einstein statistics of identical photons).  These models use different quantum phenomena in essential
       ways and they appear very different from the classical model.  Understanding the relation between these models
        will lead to a better understanding of what quantum computation is. In the present paper I show that the KLM
        model is very similar to the measurement based computation.

The KLM computation superficially resembles the gate array model. Like in the ordinary gate array model, photons
pass through a series of gates (beamsplitters, phase shifters, etc.). The main difference from the ordinary gate
array is that in the gate array, when two qubits enter a two-qubit gate, they interact. In the KLM model on the
other hand,  photons do not interact with each other. Each photon behaves exactly as if other photons were not
present. Indeed, each gate - a beamsplitter or a phase-shifter - is an optical linear element.  But how is it
possible that a computer in which the qubits are non-interacting can be as powerful as one in which they do
interact? The main issue that concerns us here is to understand where the power of computation of the KLM model
comes from.

An essential element in quantum computation (at least in algorithms that allow exponential speed-up) is entanglement
\cite{LindenPopescu} \cite{Linden Jozsa}.  Following entanglement through the computation should be illuminating.

In the gate array model entanglement is produced throughout all the stages of the computation. Indeed, the qubits proceed
from one gate to another, and at each gate that acts on two or more qubits, the qubits could get entangled.  In KLM
computation, although the photons do not interact at the gates, entanglement is still produced via peculiar interference
effects due to the identity of the photons. It is this process, and its relation with the overall computation that we
aim here to understand better.

An interesting computational model is the one-way measurement based computation. In this model entangling is
completely separated from all the other aspects of the computation. The computation starts by preparing the
qubits in a standard entangled state, independent of the particular computation that has to be performed.
Preparing this state requires (quantum) interactions between the qubits. After this stage, the qubits no longer
have quantum interactions with each other, and thus no more entangling operations are performed. Each qubit then
evolves independently \footnote {In the original model this evolution is trivial: the hamiltonian is zero}
until, at some appropriate moment,  it is subjected to a measurement.) The computation is effectively realized
by a sequence of measurements. Which qubits are measured, and what exactly each measurement is, depends on the
problem to be solved and also on the results obtained at the preceding measurements (so called ``feed-forward").

In what follows I will show that KLM computation can be easily viewed as a measurement based computation, in which
entanglement is provided at the beginning (in a standard form) and then the qubits, physically embodied by {\it
non-identical} particles,  evolve without quantum interaction.

Consider first a basic element in the KLM computation, namely a 50\%-50\% beamsplitter on which two photons
impinge simultaneously, one on each side (fig 1).
\begin{figure}[h]
\epsfig{file=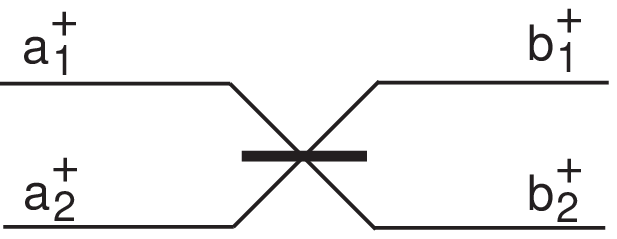} \caption {}
\end{figure}

Let $a^{\dagger}_1$ and $a^{\dagger}_2$ denote the incoming modes and $b^{\dagger}_1$ and $b^{\dagger}_2$ denote
the outgoing modes. Suppose first that a single photon impinges on the beamsplitter, in mode $a^{\dagger}_1$ .
The standard evolution is \beq a^{\dagger}_1|0>\rightarrow {1\over{\sqrt
2}}(b^{\dagger}_1+b^{\dagger}_2)|0>.\label{mode1}\eeq On the other hand, when a single photon in mode
$a^{\dagger}_2$ impinges on the beamsplitter, the evolution is \beq a^{\dagger}_2|0>\rightarrow {1\over{\sqrt
2}}(b^{\dagger}_1-b^{\dagger}_2)|0>.\label{mode2}\eeq Now, when two photons impinge together on the
beamsplitter, the evolution is \beq a^{\dagger}_2 a^{\dagger}_1|0>\rightarrow {1\over{\sqrt
2}}(b^{\dagger}_1-b^{\dagger}_2)|{1\over{\sqrt 2}} (b^{\dagger}_1+b^{\dagger}_2)|0>\nonumber\eeq \beq={1\over
2}\bigl((b^{\dagger}_1)^2-(b^{\dagger}_2)^2\bigr)|0>.\label{mandeldip}\eeq Hence, the photons emerging from the
beamsplitter are correlated, in a superposition of both being in mode $b^{\dagger}_1$ or both in mode
$b^{\dagger}_2$, that is, both leaving the beamsplitter in the same direction. (In quantum optics this effect is
known as the Hong-Ou-Mandel dip\cite{mandel}).

The situation above was analyzed in second quantization. Consider now the same situation described in first
quantization. Let $|1>$ and $|2>$ represent wave-functions of a photon impinging on the two sides  of the
beamsplitter and  $|1' >$ and $|2' >$ represent outgoing states. Then the evolutions representing a single
photon impinging on a beamsplitter,  (\ref {mode1}) and (\ref{mode2}), become
\beq |1 >\rightarrow {1\over{\sqrt 2}}(|1' >+|2' >)\label{mode1firstq}\eeq
\beq |2 >\rightarrow {1\over{\sqrt 2}}(|1' >-|2' >).\label{mode2firstq}\eeq
When two photons are impinging on the beamsplitter simultaneously, one on each side, the initial state, in first
quantization is the symmetrised state

\beq  {1\over{\sqrt 2}}(| 1>_1|2 >_2+|2>_1|1>_2)\label{initialstatefirstq}\eeq
where the indexes 1 and 2 denote the first and second photon respectively. (Following, for notation simplicity I will
drop the indexes denoting the photons; the photons being indexed by the order of the kets.) The evolution induced by
the beamsplitter is then

\beq  {1\over{\sqrt 2}}(| 1>|2 >+|2>|1>)\rightarrow \nonumber \eeq
\beq  {1\over{\sqrt 2}}\bigl( {1\over{\sqrt 2}}(|1' >+|2' >){1\over{\sqrt 2}}(|1' >-|2' >)+\nonumber\eeq
\beq {1\over{\sqrt 2}}(|1' >-|2' >){1\over{\sqrt 2}}(|1' >+|2' >)\bigr)=\nonumber \eeq \beq{1\over{\sqrt 2}}
(|1'  >|1 '>-|2' >|2' >).\label{mandeldipfirstq}\eeq

In first  quantization is clear that there is no interaction and that the entanglement of the final state is simply a
result of the entanglement present in the initial state.

As a further step, we don't actually need to consider the two photons to be identical. As long as they start in the
entangled symmetrized initial state (\ref {initialstatefirstq}) and they evolve without interaction, each according to
the evolutions (\ref {mode1}) and (\ref{mode2}), they will evolve into the entangled symmetrized state (\ref
{mandeldipfirstq}).

Finally, let's  go one more step further. Since there is no interaction, there is actually no need for the two
non-identical photons to both impinge on the same beamsplitter.  So consider two beamsplitters and two
non-identical photons, the first photon impinging on the first beamsplitter and the second photon on the second
beamsplitter (fig. 2).  When referring  to the first photon let $|1>$ and $|2>$ denote states impinging on the
two sides of the first beamsplitter while when referring to photon 2 the states  $|1>$ and $|2>$ denote states
impinging on the two sides of the second beamsplitter. Similarly, the notation $|1' >$ and $|2' >$ denotes
outgoing states from the first beamsplitter when referring to the first photon, and outgoing states from the
second beamsplitter when referring to the second photon. Now, although each photon impinges on a separate
beamsplitter, it is still the case that if the two photons start in the state (\ref {initialstatefirstq}) they
end up in the final state (\ref{mandeldipfirstq}).  It is now clear that the entanglement in the final state are
a direct result of the entanglement existing in the initial state. Indeed, since the photons undergo unitary
evolutions and they do not interact with each other, the entanglement is conserved.
\begin{figure}[h]
\epsfig{file=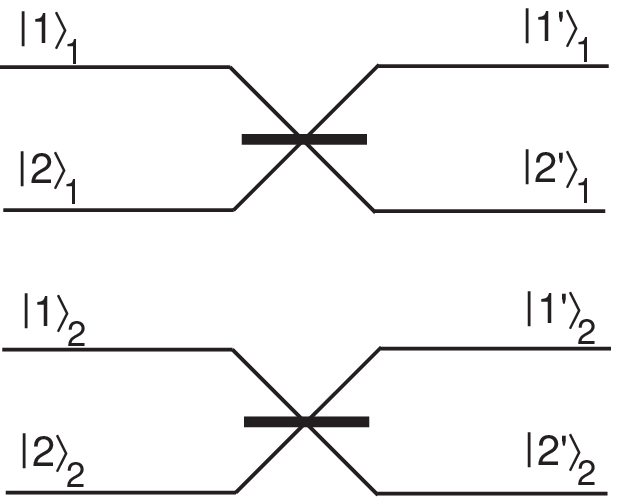} \caption {}
\end{figure}

With the above simple example understood, we can now approach the general KLM computer. Rather than discussing
KLM computation gate by gate, let us consider an entire KLM computer. A KLM computer is in effect a large
interferometer, composed of beamsplitters and phase shifters. In addition, the computer contains photon
detectors and conditional phase shifters -  phase shifters that act or not depending on a classical input,
which, in its turn, may depend on the outcome(s) of a detector(s).

The detectors used in KLM computation are ``single photon" detectors.  They are supposed to give a click
whenever one or more photons are presents. Now, there are different ways in which such detectors can work. They
might make a coherent measurement on the subspace of one or more photons, or they may decohere these subspaces
even though the detector doesn't indicate the precise number of photons. However, for our purposes it doesn't
really matter how the detectors actually work. Indeed, the photons that impinge on a detector are absorbed and
no longer used in the interferometer. Then, by unitarity, the information about how many photons were absorbed
will always be present and could, in principle, be recovered at a later time. So as far as the photons that
remain in the interferometer are concerned, everything happens as if the detectors decohere the subspaces
corresponding to different photon numbers. In effect, we can assume that the detectors in the original KLM
scheme are actually photon number detectors, that is, they measure the number of photons; making this assumption
cannot lead to a weaker scheme.

The input into the KLM  computer is always given in the form of a state of $N$ photons,  in a direct product
state of one photon in each of $N$ input modes, i.e. \beq |\Psi_{in}>=a^{\dagger}_N...a^{\dagger}_2a^{\dagger}_1
|0>.\label{generalinitialstate}\eeq Note that in the $N$ photons are included also all the photons that play the
role of ancillas in the implementations of the different gates.

The main result of this paper is that we can map any KLM computer on an equivalent circuit in the following way,
illustrated in fig. 3.

We  map any KLM interferometer onto $n$ independent interferometers, each identical to the original one, with a single
(non-identical) photon in each interferometer. Each interferometer contains also detectors and controlled phase shifters
as the original KLM interferometer. The equivalent detectors from each interferometer  are all connected, and whenever
one of them clicks, all the corresponding controlled phase shifters in all the interferometers are activated. (Actually,
 there is no need to activate the phase shifter in the interferometer  whose detector clicked, since there is no photon
 it can act upon - indeed the (single) photon in that particular interferometer has already been absorbed by the
 detector)

Finally, the initial state is the ``permanent" state of the $n$ photons, i.e.
\beq |\Psi>= S |1>|2>...|n>,\label{permanent}\eeq
where the operator $S$ means total symmetrizarion. (We use the term ``permanent"  since the state can be obtained
from the Slater type matrix
\begin{equation}
\left( \begin{array}{cccc} \ket{1} & \ket{2} & \cdots & \ket{n} \\  \ket{1} & \ket{2} & \cdots & \ket{n} \\
\vdots & \vdots & \cdots & \vdots \\ \ket{1} & \ket{2} & \cdots & \ket{n}  \end{array} \right) \end{equation} by
calculating the permanent, instead of the determinant as needed for the totally anti-symmetric state. The
permanent
 is identical to the determinant but has +1 factors in front of each term, while the determinant has +1 in front of
 some of the terms and -1 in front of others.)

First of all, if the original interferometer would only have linear elements but not detectors and controlled
phase shifters, than it is a simple exercise to generalize the analysis of the Hong-Ou-Mandel dip made before
and to prove the validity of the proposed mapping.  To complete the analysis of the KLM computation we need now
to discuss the detectors and conditional phase-shifters.  Clearly, having one photon number detector in each
interferometer makes an overall photon number detector over the $n$ interferometers. (Note that this is the case
also if each detector can only differentiate between zero and more than one photons - each interferometer only
contains a single photon, so what the detectors do when more photons are present is irrelevant.) Having an
effective photon number detector rather than a detector that can distinguish only the zero photon subspace from
the subspace of one and more photons as in the original KLM scheme is however not a problem, as discussed
before.

There is one further delicate aspect related to photon detection. Suppose that in the original KLM scheme there is
one photon in the mode that is detected. Then that photon is absorbed, and this is the end of the story. On the other
hand, in first quantization  we don't know which of the identical photons is detected.  This seems to lead to problems
 because in our scheme we do know which photon is detected - we just look at which detector clicks. So the
 correspondence between the first quantization picture (that was the starting point for our map) and the
 multi-intreferometers scheme seems to break down. Indeed, the initial symmetrization is now lost. However, this is not
 actually a problem because the detected photon is absorbed and no longer plays any role while the remaining photons
 continue to be in a symmetrized state.

Finally, our mapping also requires that once a detector clicks, then it activates all the appropriate equivalent
controlled phase shifters in the other interferometers. The reason is that in the original KLM scheme, once a detector
clicks, it activates a phase shifter that acts on all photons that populate the particular mode on which the phase
shifter acts.

\begin{figure}[h]
\epsfig{file=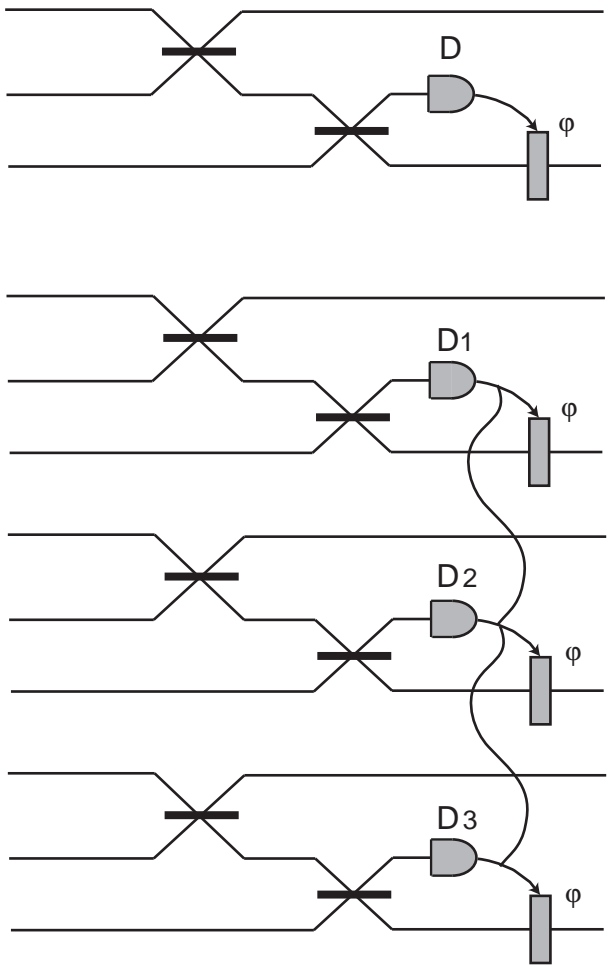} \caption{$D$, $D_1$, $D_2$ and $D_3$ are detectors and $\varphi$ are phase shifters
controlled by the detectors. In (a) there are three identical photons, one in each input mode. In (b) there is
one single photon in each interferometer, the initial state being the permanent state. The detectors are
connected so that when one of them fires all the phase shifters are activated. Note that the interferomener
illustrated here doesn't actually represent any particular KLM computation, but has all its defining elements.}
\end{figure}

Let us now analyze our results. The whole reason for mapping the original KLM computer onto the $n$ independent
interferometers model is that the later is actually a  measurement based one-way computer.  The $n$  non-identical
photons, start in a standard entangled state, (the permanent state) and each photon evolves independently.
The computation is then driven by measurements performed onto the individual photons and feed-forward (by the result
of each  measurement producing controlled phase-shifts of the phases of other photons).  But, and this is a most
important point, it is not only our  $n$ interferometers model that is an one-way measurement- based computation;
the same is true for the original KLM model as described in first quantization. This gives an explanation of why the
KLM scheme can work despite the fact that the photons do not interact: they don't have to. The scheme is is a measurement
 based computation in which the entanglement has already been supplied from the very beginning (via the symmetrization
 of the initial state of the $n$ identical photons.

Finally, a corollary of the present analysis is that the permanent state of $n$ quantum systems, each described
by an $n$ dimensional Hilbert space,  is universal for quantum computation.  Indeed, we know that the KLM
computation is universal, and the above analysis showed that the permanent state allows us to implement KLM
computation. The permanent state therefore joins the cluster state \cite {oneway} as another state universal for
quantum computation.

\begin{acknowledgments}
{\bf Acknowledgments}  The author would like to acknowledge very useful discussions with A. Short and N. Yoran.
This work was supported by the UK EPSRC grant GR/527405/01 and the UK EPSRC ``QIP IRC"
project.\end{acknowledgments}


\begin{thebibliography}{11}

\bibitem{adiabatic}E. Farhi, J. Goldstone, S. Gutmann, and M. Sipser, quant-ph/0001106.
\bibitem{topological} A. Kitaev quant-ph/9707021.
\bibitem{oneway}R. Raussendorf and H. J. Briegel, Phys. Rev. Lett. 86, (2001), 5188.
\bibitem{klm} E. Knill, R. Laflamme and G. J. Milburn, Nature 409, (2001), 46.
\bibitem{LindenPopescu} N. Linden and S. Popescu, Phys.Rev.Lett 87, 047901.
\bibitem{Linden Jozsa} R. Jozsa and N. Linden, Proc. Roy. Soc. 459, (2003), 2011.
\bibitem{mandel} Hong C K, Ou Z Y and Mandel L  Phys. Rev. Lett. 59, (1987), 2044.


\end{thebibliography}
\end{document}